\documentclass{optica-article}

\journal{opticajournal} 

\articletype{Research Article}

\usepackage{lineno}

\begin{document}

\title{Variational MineGAN: A Data-efficient Knowledge Transfer Architecture for Generative AI-assisted Design of Nanophotonic Structures}

\author{Shahriar Tarvir Nushin,\authormark{1} Shadman Shahriar Sharar,\authormark{1,*} and Farhan Ishraque Zahin\authormark{2}}

\address{\authormark{1}Department of Electrical and Electronics Engineering, Islamic University of Technology, Dhaka, Bangladesh\\
\authormark{2}School of Electrical and Computer Engineering, Purdue University, West Lafayette, Indiana 47907-1285, United States}

\email{\authormark{*}shadmanshahriar5@iut-dhaka.edu} 


\begin{abstract*} 
Using deep learning to design nanophotonic devices has recently been an active research area, with Generative Adversarial Networks (GANs) being a popular choice alongside autoencoder-based methods. However, both generally require large datasets and computational power, creating limitations in data-scarce scenarios. Fine-tuning GANs on limited data often cause mode collapse and overfitting, reducing generalizability and model effectiveness. To address this, we introduce Variational MineGAN, an optimized architecture that enhances data efficiency by reducing overfitting. Experimental results demonstrate a Frechet Inception Distance (FID) of 52.14 and an Inception Score (IS) of 3.59, enabling high-quality design generation and accurate spectral response estimation, improving nanophotonics design exploration.
\end{abstract*}

\section{Introduction}
Understanding the impact of intricate geometrical shapes on the properties of micro or nanophotonic devices has been among the key challenges in photonics research\cite{PhysRevB.102.035404, Qiao:18}. The conventional design approach typically begins with an educated guess about the structure's shape, followed by the use of Finite-Difference Time-Domain (FDTD) or Finite Element Method (FEM) solver software to solve Maxwell’s partial differential equations \cite{HonglingRao992583,rahman2013finite}to obtain the spectral response of the structure. The design parameters are then tweaked to optimize performance and iteratively evaluated through computationally expensive simulations. This trial-and-error method is both time and resource-intensive \cite{KUANGCHEN1999171, Duffy2004, Zauderer2011}.

To address these limitations, different methods based on deep learning (DL) have recently been employed for the design of nanophotonic structures. Discriminative deep learning models have been utilized for both forward and inverse electrodynamics problems \cite{nadell2019deep,peurifoy2018nanophotonic,ma2018deep}. Additionally, generative deep learning models such as Generative Adversarial Networks (GANs)\cite{liu2018generative} and Variational Autoencoders (VAEs)\cite{ma2019probabilistic} have been utilized for design optimization. Different variations of generative model architectures have been proven to be effective in solving these optimization problems. Kudyshev et al. used an adversarial autoencoder for topology optimization of thermal emitters \cite{kudyshev2020machinelearningassistedglobal}. Panisilvam et al. proposed the use of asymmetric CycleGAN for the inverse design of photonic devices \cite{Karras2017}. These generative models excel at learning the underlying distribution of data and generating realistic images, making them a popular choice for optimizing nanophotonic structures based on desired spectral responses. However, they typically require large datasets and extensive training time, which can be a significant limitation.\cite{miyato2018spectral,miyato2018cgans}

To mitigate these challenges, the concept of ``knowledge transfer" has been applied in generative model training. This approach leverages pre-trained generative models trained on large datasets for a longer time and adapts them to new, smaller datasets. Weng et al. demonstrated the effectiveness of using knowledge transfer to fine-tune the pre-trained GAN model on limited data \cite{wang2018transferringgansgeneratingimages}. However, Noguchi and Harada addressed the issue of mode collapse associated with this approach and proposed freezing certain parameters of the generator\cite{noguchi2019imagegenerationsmalldatasets}. While this approach is less prone to overfitting, it significantly limits the model's flexibility during fine-tuning\cite{Wang2019}.

To further address the shortcomings of transfer learning of the GAN model on limited data, Wang et al. introduced the ``MineGAN" approach \cite{Wang2019}. This method employs a miner network that transforms the input to a new distribution of target data, which facilitates the generator model in generating images from the new target domain.

In this work, we introduce ``Variational MineGAN," an extended version of MineGAN. Our approach modifies the miner network to directly learn the properties of the target data distribution by estimating its mean and variance instead of transforming the input to follow that distribution. This allows better understanding and control over the latent distribution space and reduces computational cost as well since it bypasses the use of the miner network during inference, streamlining the image generation process.

\section{Methodology}
 
\subsection{GAN Formulation}
A Generative Adversarial Network (GAN) comprises a pair of networks: the Generator and the Discriminator. The discriminator's job is to differentiate between actual and generated fake samples, while the generator's job is to map random noise samples into images. Goodfellow et al. presented the initial GAN formulation \cite{Goodfellow2014}, which seeks to solve a two-player min-max problem by reaching a Nash equilibrium, which is mathematically expressed as:
\begin{equation}
    \min_G \max_D \mathbb{E}_{x \sim p_{\text{data}}(x)} [\log D(x)] + \mathbb{E}_{z \sim p(z)} [\log(1 - D(G(z)))],
\end{equation} in which \(z\epsilon R^{d_z}\) is a latent variable obtained from a distribution \(p(z)\) typically a normal distribution with identity covariance \(\mathcal{N}(0,I) \) or uniform distribution \(\mathcal{U}[-1,1]\). On the other hand, \(x\sim p_{\textnormal{data}}(x)\) corresponds to samples drawn from the real data distribution. In this work, for better training stability, Wasserstein GAN-Gradient Penatly (WGAN-GP) \cite{Gulrajani2017} loss has been used for the discriminator, which utilizes the Wasserstein Loss \cite{arjovsky2017wassersteingan} alongside a penalty term added to it, instead of Binary Cross Entropy (BCE) Loss. The discriminator's modified loss function then takes on the following form:

\begin{equation}
  L_D = \mathbb{E}_{{z} \sim \mathbb{P}(z)} [D(G(z))] - \mathbb{E}_{x \sim \mathbb{P}_{data}(x)} [D(x)] + \lambda \mathbb{E}_{\hat{x} \sim \mathbb{P}_{\hat{x}}} \left[ \left( \|\nabla_{\hat{x}} D(\hat{x})\|_2 - 1 \right)^2 \right].
\end{equation} Here, $\hat{x}$ represents samples that are interpolated between real data samples $x_{real} \sim {P_{data}(x)}$ and generated fake data samples, $x_{fake}= G(z)$ where $z\sim p(z)$. The term $\lambda$ is a regularization coefficient that controls the strength of the gradient penalty. The interpolation is done by
\begin{equation}
    \hat{x} =\alpha x_{real} + (1-\alpha)x_{fake},
\end{equation} where $\alpha$ is sampled from a uniform distribution $\mathcal{U}[-1,1]$ \cite{Gulrajani2017}.

\begin{figure}[ht!]
\centering\includegraphics[width=\columnwidth]{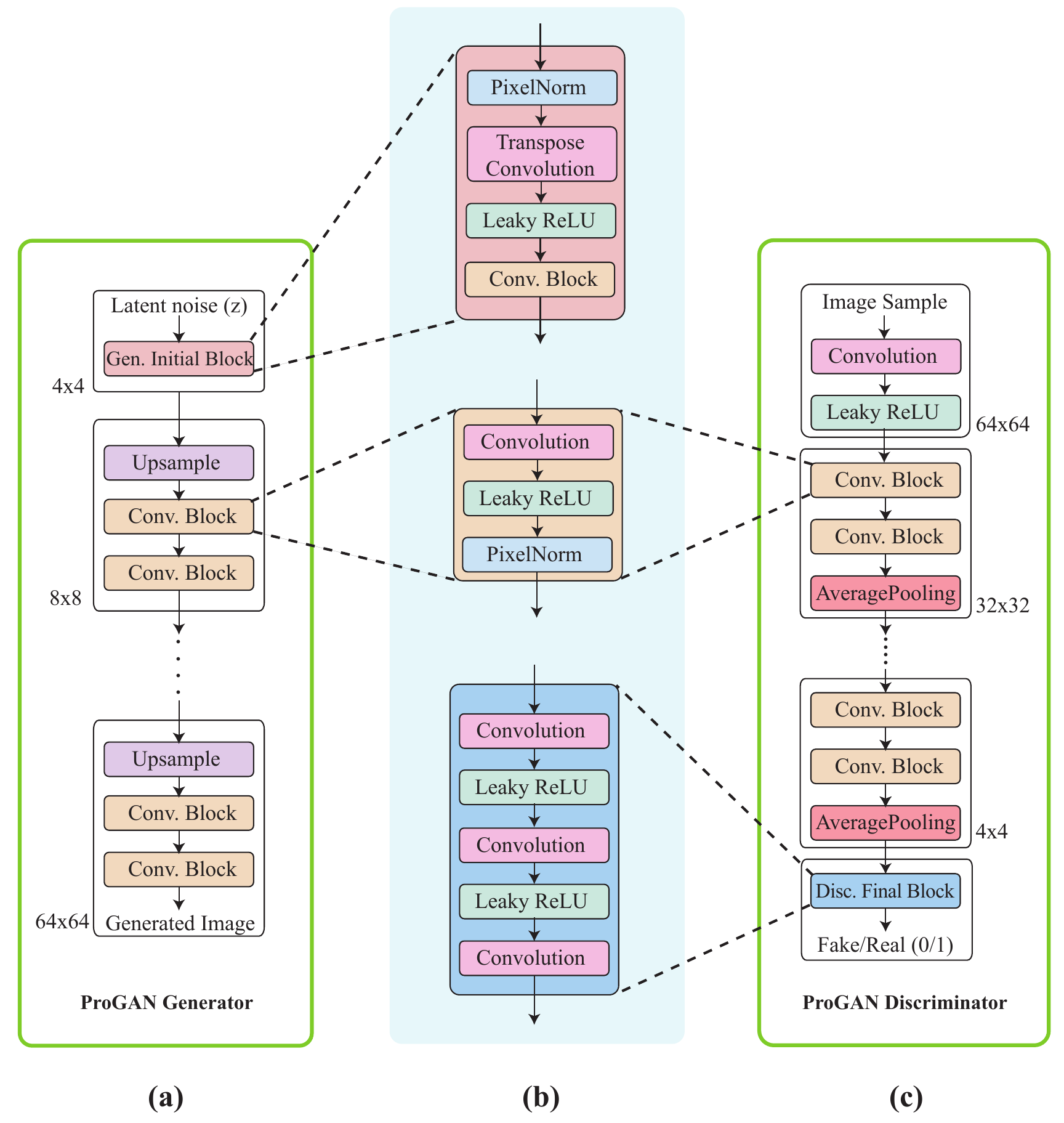}
\caption{ Overview of the ProGAN architecture. (a) ProGAN Generator: progressively generates 4$\times$4 to 64$\times$64 images from input noise. (b) Building blocks used in generator and discriminator. (c) ProGAN Discriminator: makes a real/fake prediction from input image samples.}
\label{fig:proGAN_arch}
\end{figure}

In this work, for both the generator and discriminator, we follow the Progressive GAN (Pro-GAN) architecture \cite{Karras2017}, where the training begins with images of lower resolution, and the resolution is progressively enhanced by including additional layers to the network. Although the original Pro-GAN paper starts training at 4$\times$4 and goes up to 1024$\times$1024 resolution, due to dataset constraints, we train up to 64$\times$64 resolution images. The latent noise dimension used in this work is 256, and the noise vector is drawn from \(\mathcal{N}(0,I) \). The generator and discriminator network architecture has been illustrated in Figure \ref{fig:proGAN_arch}. The convolution layers inside the ``Convolution Block" (Conv. Block) employ a kernel size of 3, stride of 1, and padding of 1. The generator uses 256 filters for convolution blocks at 4$\times$4 to 32$\times$32 image sizes and 128 filters for the 64$\times$64 size. In contrast, the discriminator follows the opposite filter setup. In the discriminator's final block, the three convolution layers use kernel sizes of 3, 4, and 1 with no padding. The output of the last convolution layer skips activation, as WGAN is used. The value of $\lambda$ for WGAN-GP is set to 10. The core of the generator and the discriminator network are mirror images of each other, and they both always grow in harmony. This ProGAN generator and discriminator are later on used as pre-trained models for both MineGAN and Variational MineGAN.

\subsection{MineGAN}

\begin{figure}[ht!]
\centering\includegraphics[width=\columnwidth]{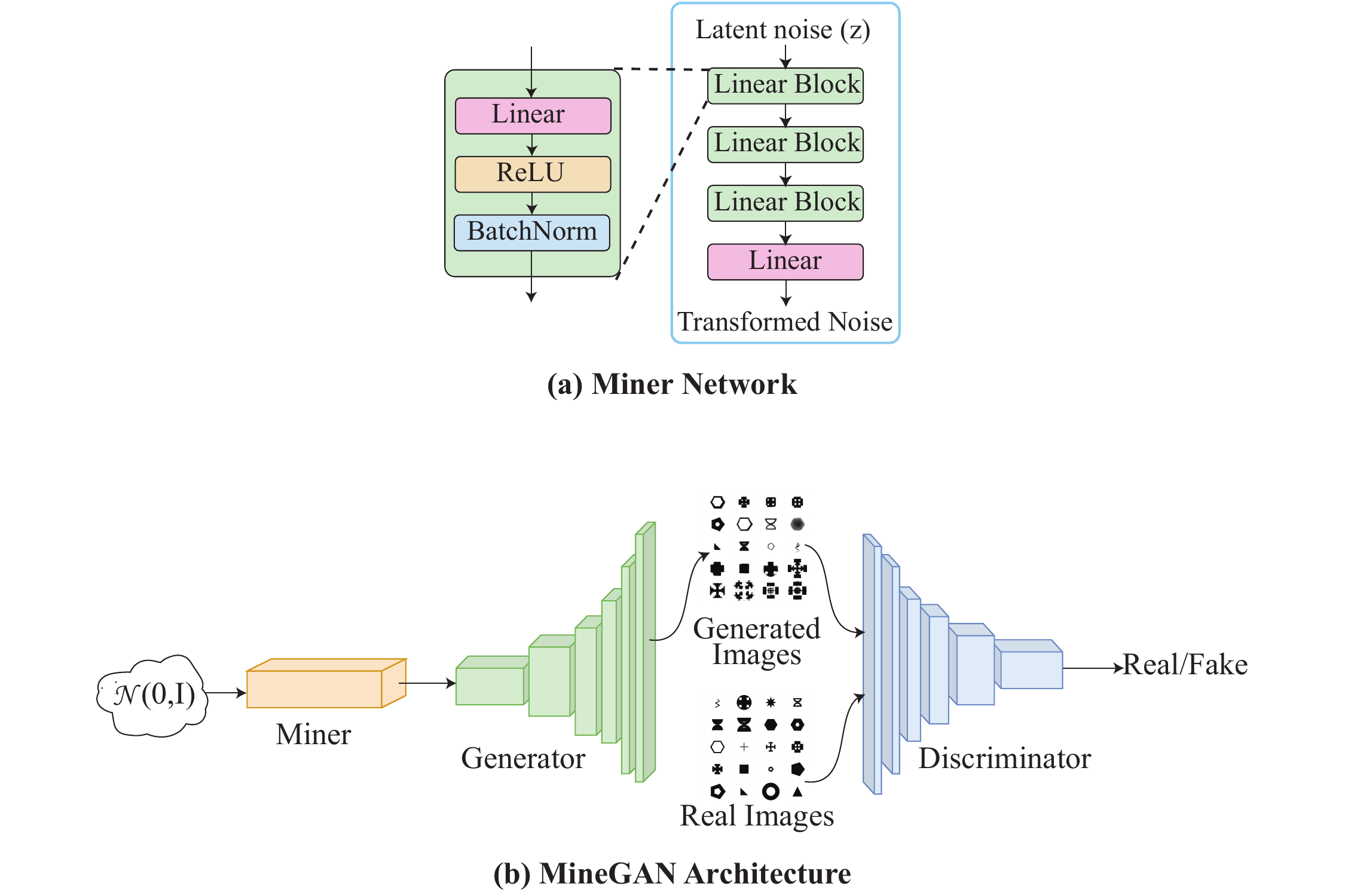}
\caption{(a) Architecture of the miner network, which is composed of multiple linear blocks, transforming input noise before passing it to the generator. (b) The overall model architecture of MineGAN.}
\label{fig:MineGAN}
\end{figure}

MineGAN employs a miner network that steers the input noise to the most promising regions of the latent space with respect to the target data distribution \(D^{T}\)\cite{Wang2019}. The miner network is a compact neural network composed of a sequence of linear layers, each preceding ReLU activation and batch normalization except for the last layer. Each linear layer has a dimension of 256. Figure \ref{fig:MineGAN} (a) illustrates the detailed configuration of the linear blocks used in the miner network and the configuration of the miner network itself used in our implementation. 

Let our Generator and Discriminator be trained on prior data distribution \(p_{\textnormal{data}}\). We want to fine-tune the pre-trained model on new target data distribution \(p_{\textnormal{data}}^{T}\). The miner functions as an intermediary between the input noise vector and the generator. The input passes through the miner network at first, where it is transformed before entering the generator, in contrast to traditional GANs, where the noise flows directly to the generator. The discriminator part remains the same as usual, taking real and generated fake images as input and attempting to classify them. Figure \ref{fig:MineGAN} (b) demonstrates the overall architecture of the MineGAN network.

MineGAN training takes place in two stages. Initially, the generator weights are kept fixed, and only the miner is trained. In this stage, the miner network learns to transform the input noise \(z\sim p(z)\) in such a manner that it follows the new target data distribution. The generator is no longer fixed during the second stage of training, effectively fine-tuning the entire architecture. However, since the miner network is composed of significantly less amount of parameters compared to the GAN network, it is, therefore, less susceptible to overfitting\cite{Wang2019}. It narrows the latent distribution to align more with the target, simplifying fine-tuning by providing a more consistent training signal, and as a result, the generator weights do not require to go through significant updates, thus reducing the likelihood of overfitting. The employed loss functions for the Generator and the Discriminator are
\begin{equation} L_G = -\mathbb{E}_{z \sim p(z)} [D(G(M(z)))],
\end{equation}
and
\begin{equation}
        L_D = \mathbb{E}_{z \sim p(z)} [D(G(M(z)))] - \mathbb{E}_{x \sim p_T(x)} [D(x)]
\end{equation}
respectively.

\subsection{Variational MineGAN}
\begin{figure}[ht!]
\centering\includegraphics[width=\columnwidth]{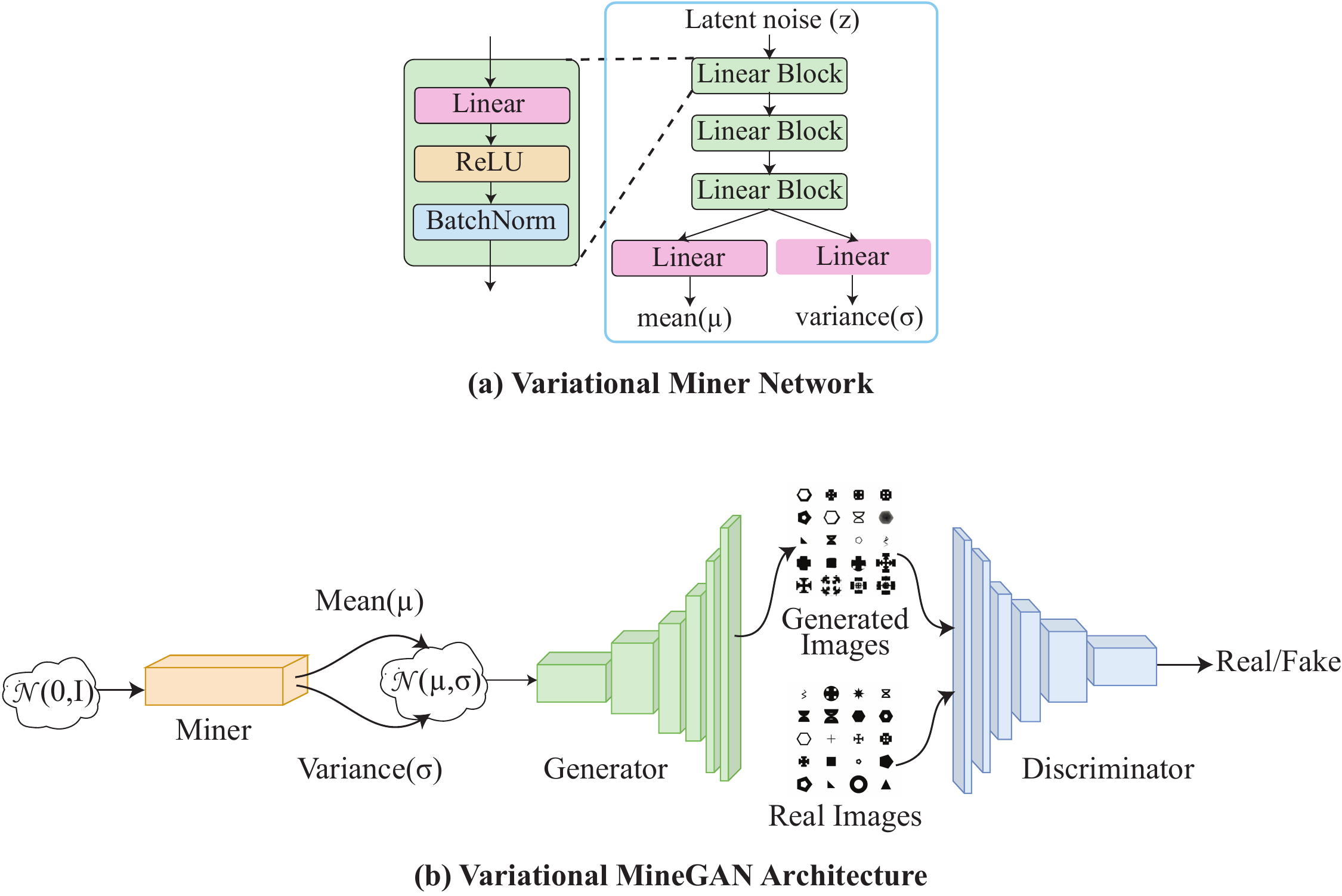}
\caption{(a)Architecture of the variational miner network, where input noise is processed through a series of linear blocks to output the mean ($\mu$) and variance ($\sigma$) of the target distribution.(b) The overall model architecture of Variational MineGAN.}
\label{fig:VarMineGAN}
\end{figure}

In this work, we extend the MineGAN framework by introducing variational MineGAN. Instead of directly transforming the noise as in MineGAN, our approach uses a variational miner network. Similar to the miner network, the variational miner is also a compact neural network; however, it has two outputs corresponding to the mean and the variance of the target distribution. Each linear layer has a dimension of 256. Figure \ref{fig:VarMineGAN} (a) illustrates the details of the variational miner network.

In Variational MineGAN, the input noise \(z\sim p(z)\) goes through the miner network, and the network gives us the parameters of a new distribution that aligns much more closely with the target distribution. Figure \ref{fig:VarMineGAN} (b) illustrates the overall architecture of the Variational MineGAN network. Noise samples are then taken from the new distribution and passed on to the generator. This modification allows for better exploration of the latent space, which ensures that the generated image samples are more representative of the target data distribution while reducing the risk of overfitting. The variational approach also adds a form of regularization, promoting better generalization of the model to unseen data.

Another significant advantage of the Variational MineGAN is the reduced computational overhead during inference. Once the mean and the variance of the target distribution have been learned by the miner, there is no need to use the miner for inference. Instead, the model uses the learned parameters to directly sample from the latent space, resulting in a more efficient image generation process. In contrast, MineGAN requires the miner to transform the noise input during inference, which adds computational cost. This transformation, while effective, operates somewhat like a black box, as we don't have direct insight into the distribution of the transformed noise. With Variational MineGAN, however, we gain a clear understanding of the new distribution space as we explicitly model the mean and variance. This transparency provides insight into how the noise is sampled and controlled, improving interpretability without sacrificing performance.

\subsection{Absorption Spectra Prediction Model}

\begin{figure}[ht!]
\centering\includegraphics[width=\columnwidth]{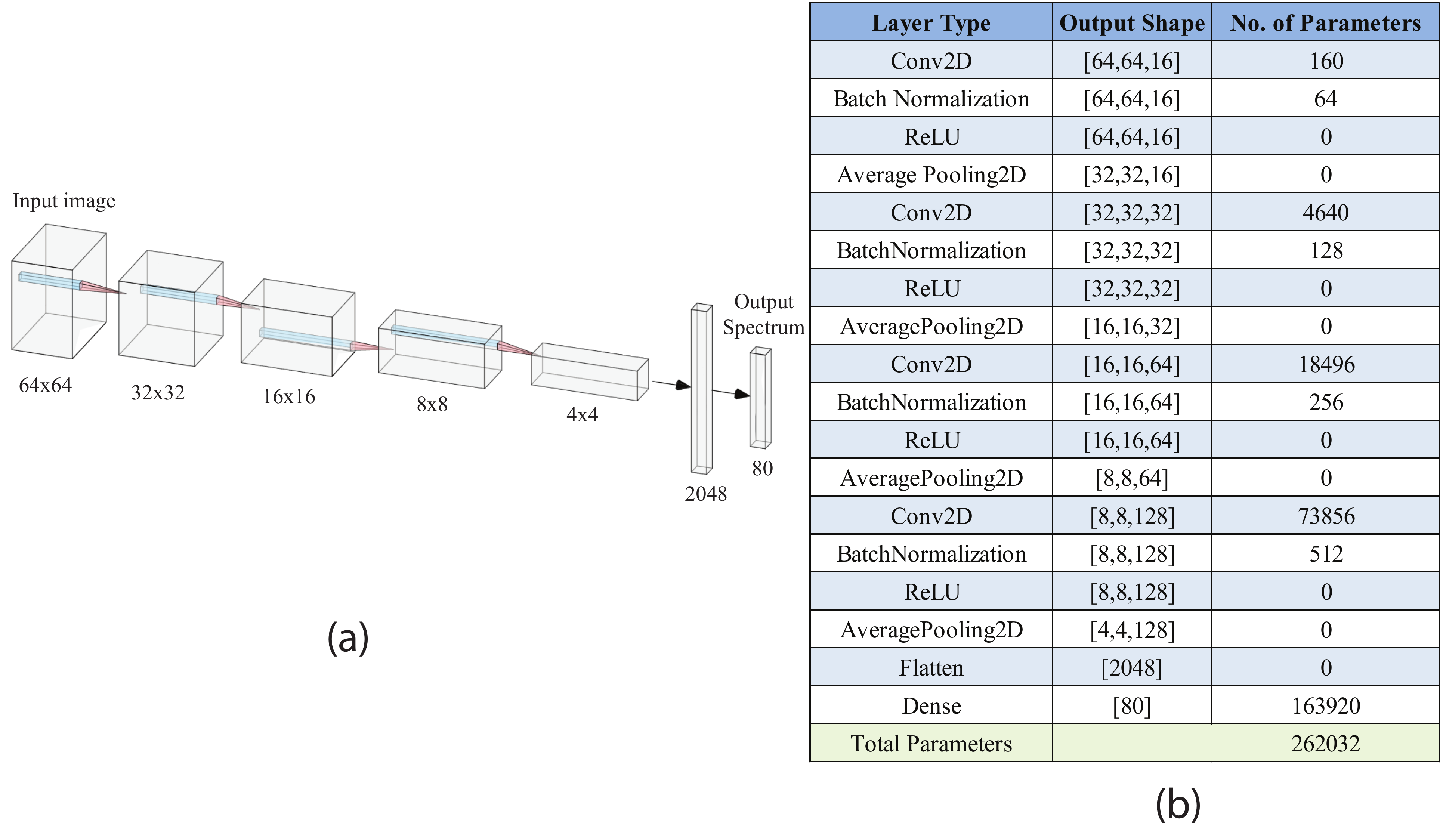}
\caption{Absorption spectra prediction overview: (a) Model architecture, (b) Layer details.}
\label{fig:Predict_archi}
\end{figure}

In addition to training the generative models, a separate predictive model was also trained to estimate the absorption spectra based on the design images generated. A Convolutional Neural Network (CNN) makes up this model that takes a 64\(\times\)64 pixel grayscale design image as input, and 80 spectral values corresponding to wavelengths ranging from 4$\mu$m to 12$\mu$m come out as output. The output of this model is passed through a smoothing filter with a window size of 5 to obtain a smooth spectral response. The architecture of this model has been illustrated in Figure \ref{fig:Predict_archi}. This model allows us to efficiently select designs that meet specific performance parameters before conducting more resource-intensive simulations for further optimization.

\subsection{Dataset Details}
\label{sec:dataset}
To pre-train our ProGAN model, we utilized a publicly available dataset from Raman Lab of UCLA comprising 18,770 images of metamaterial absorbers \cite{Yeung2021}. Lumerical software's 3D Finite-Difference Time-Domain (FDTD) simulations were used to create these data. The dataset features seven distinct shapes, each with geometric parameter variations. These structures had 3.2\(\times\)3.2 \(\mu m^2\) as their unit cell dimensions. In order to simulate periodic structures, perfectly matched layers (PMLs) were employed above and below the plane of interest, and periodic boundary conditions were implemented on the side boundaries. A source plane was injected parallel to and at a distance from the plane of interest. Reflectance and transmission spectra were calculated using the above-mentioned simulation arrangement.  

For the fine-tuning stage with MineGAN and Variational MineGAN, we constructed a smaller target dataset of 1000 images by combining images from another dataset from Raman lab \cite{Yeung2020}  with an additional 200 images of designs that we simulated, which involved variations of different geometric structures. The physical configuration of the designs from this dataset is similar to the one described previously. This dataset also contains the absorption spectra within the range of 4$\mu$m to 12$\mu$m wavelength on a separate file. This dataset was used to train the MineGAN and Variational MineGAN models, as well as the spectra prediction model, incorporating a diverse range of geometric shapes and sizes. This approach ensured that the target dataset introduced new variations that the pre-trained model had not encountered before.

\subsection{Training Details}
\subsubsection{ProGAN pre-training}
For the pre-trained network, a ProGAN architecture was trained on the multiclass dataset of Raman Lab \cite{Yeung2021}. Training started at 4\(\times\)4 resolution, then 8x8, 16\(\times\)16 and gradually up to 64\(\times\)64 pixels. The models were trained for 30 epochs at each resolution. PixelNorm was applied after each convolution layer, and minibatch standard deviation information was integrated into the channel dimension of the discriminator. The learning rate was configured to 0.001 using the Adam optimizer, Lambda-GP was set to 10, and one discriminator iteration was performed per generator update.

\subsubsection{MineGAN and Variational MineGAN Training}
Both MineGAN and Variational MineGAN networks were trained on the smaller target dataset of 1000 images to create sensor images that were not a part of the initial dataset on which the ProGAN model was trained first. Both the generator and the discriminator were initialized with pre-trained weights. The training took place in two stages. For stage 1 of training, the miner was trained for 1000 epochs, keeping the generator weights frozen. The generator was then trained for 300 epochs, and there was a relatively lower learning rate of 0.01 during the second stage of training.

\subsubsection{Absorption Spectra Prediction Model Training}
The dataset for this training consisted of real images paired with their corresponding absorption spectra, which served as ground truth. The network was trained using mean squared error (MSE) as the loss function and Adam optimizer with a learning rate of 0.001. The model underwent training with a batch size of 32 across 250 epochs. The dataset was divided into training and testing portions in a ratio of 80 to 20.

\section{Results and Discussion}

We conducted a comprehensive evaluation of four different generative approaches: training from scratch (ProGAN trained directly on the target dataset), Transfer GAN \cite{wang2018transferringgansgeneratingimages}, MineGAN, and our proposed Variational MineGAN. Each model was evaluated using two key performance metrics, Frechet Inception Distance (FID)\cite{heusel2018ganstrainedtimescaleupdate} and Inception Score (IS) \cite{NIPS2016_8a3363ab}. FID measures the similarity between real and generated images by comparing the distributions of feature representations. On the other hand, IS evaluates the quality and diversity of generated images by assessing how confidently the images are classified into distinct classes. A lower FID score and higher IS are desirable since they indicate that the generated images are realistic and exhibit a greater diversity of features\cite{memari2024advancinggenerativemodelevaluation,areganeuqal}. 

Table \ref{tab:results} highlights the best FID and Inception scores achieved on the 1000 image dataset across all approaches. The Variational MineGAN shows a significant improvement over the other methods, recording the lowest FID scores and the highest IS values, signifying its superior ability to generate images of superior quality that closely match the real data distribution.

\begin{table}[ht!]
    \centering
    \begin{tabular}{|c|c|c|}
        \hline
        \textbf{Approach} & \textbf{Best FID Score} & \textbf{Inception Score (IS)} \\
        \hline
        Model Trained from Scratch & 105.51 & 3.03 \\
        \hline
        Transfer GAN & 73.18 & 3.36 \\
        \hline
        MineGAN & 59.23 & 3.49 \\
        \hline
        Variational MineGAN & \textbf{52.14} & \textbf{3.59} \\
        \hline
    \end{tabular}
        \caption{Best FID and IS for different approaches on the target dataset of 1000 images.}
    \label{tab:results}
\end{table}

\begin{figure}[ht!]
\centering\includegraphics[width=\columnwidth]{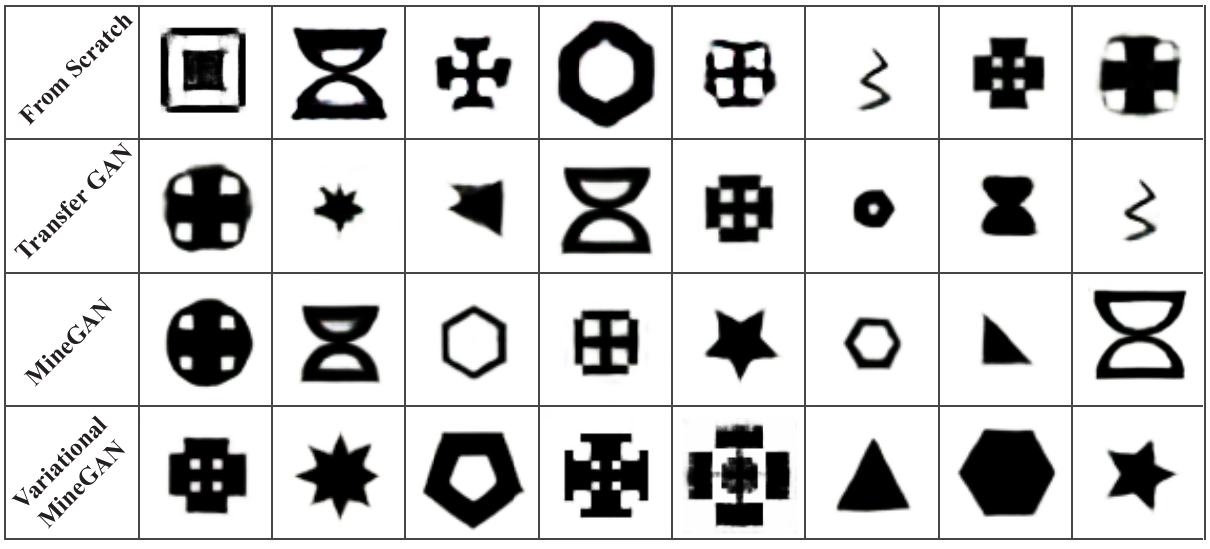}
\caption{Sample images generated by each approach.}
\label{fig:photogrid}
\end{figure}

The FID scores, in particular, show a marked difference when compared to the models trained from scratch or using Transfer GAN without a miner. The variational component not only improves the model's generalization capabilities but also reduces overfitting, which is a common challenge in GAN training, particularly with small datasets\cite{zhao2020differentiableaugmentationdataefficientgan}. Figure \ref{fig:photogrid} shows that Variational MineGAN generates clearer, more realistic images compared to other approaches. Images from models trained from scratch and Transfer GAN show more artifacts and lack sharpness, which becomes more prominent in smaller datasets.

\begin{figure}[ht!]
\centering\includegraphics[width=\columnwidth]{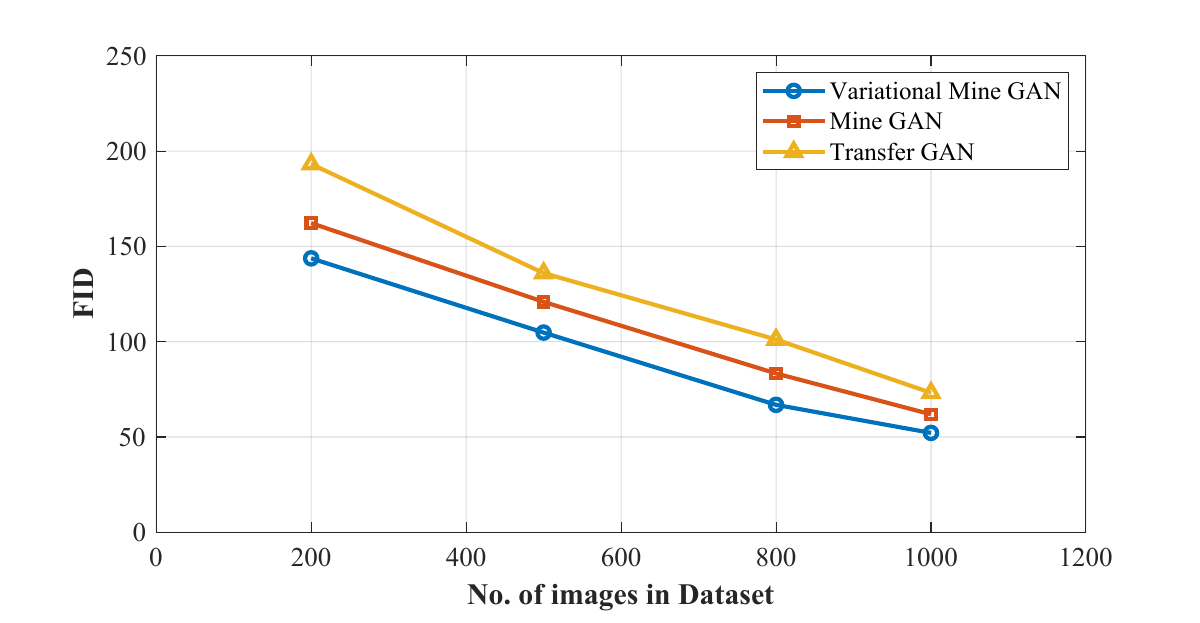}
\caption{FID score progression with increasing size of the target dataset.}
\label{fig:FID_progression}
\end{figure}

To analyze how well these models adapt to limited data, experiments were carried out with varying dataset sizes of 200, 500, 800, and 1,000 images. As shown in Figure \ref{fig:FID_progression}, the FID scores of all models improve as the dataset size increases. However, the Variational MineGAN consistently outperforms the other approaches at every dataset size, maintaining lower FID scores even when only 200 images were used.

The models trained from scratch and Transfer GAN experience more difficulty adapting to smaller datasets, leading to higher FID scores and signs of overfitting. MineGAN performs better than these methods due to the incorporation of a miner network that better aligns the latent space with the target distribution, but it still falls short of the performance demonstrated by Variational MineGAN.

\begin{figure}[ht!]
\centering\includegraphics[width=\columnwidth]{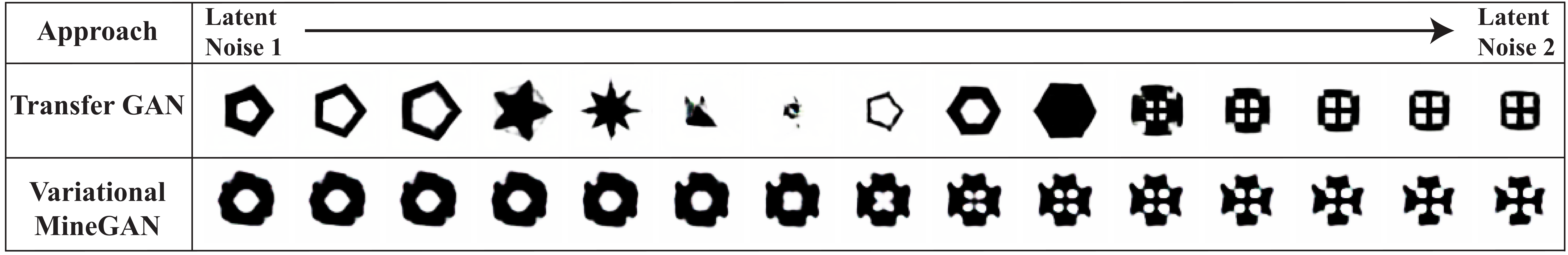}
\caption{Smooth transitions by Variational MineGAN vs. erratic ones by TransferGAN in case of interpolation between two latent noise vectors.}
\label{fig:Interpolate}
\end{figure}

To further investigate the degree of overfitting, we conducted an interpolation experiment by generating images from interpolations between two latent noise vectors. This test helps to assess how well each model captures the global structure of the data. Models that overfit tend to generate disjointed or abrupt transitions between interpolated samples, while models that generalize well will produce smooth and coherent transitions\cite{Radford2015UnsupervisedRL,arora2017generalizationequilibriumgenerativeadversarial}. As illustrated in figure \ref{fig:Interpolate}, Variational MineGAN produces smooth and continuous image transitions, demonstrating its ability to identify the underlying data distribution effectively. On the other hand, Transfer GAN produced more erratic transitions, indicating a higher level of overfitting to the training data. This observation further supports the conclusion that Variational MineGAN better generalizes to unseen data.

Moreover, in the case of variational MineGAN, the miner network is no longer required during inference once we have the information of the new distribution space, allowing us to sample noise from this new distribution and generate images directly, simplifying the process and reducing computational overhead. Table \ref{tab:generation_times} reflects the increase in time difference between MineGAn and variational MineGAN approach as we generate more images. The efficiency difference becomes more apparent when generating large batches of images. For instance, generating 10,000 images with Variational MineGAN took approximately \textbf{38.49} seconds, whereas MineGAN required \textbf{45.23} seconds. This reduction in computational demand can significantly benefit large-scale image generation tasks, making Variational MineGAN a more practical choice.

\begin{table}[ht!]
    \centering
    \begin{tabular}{|c|c|c|c|}
        \hline
        \textbf{Method} & \textbf{1000 Images}& \textbf{5000 Images} & \textbf{10000 Images} \\
        \hline
        MineGAN & 6.01 sec & 23.22 sec & 45.23 sec\\
        \hline
        Variational MineGAN &  4.58 sec & 19.78 sec & 38.49 sec \\
        \hline
    \end{tabular}
    \caption{Inference time comparison between MineGAN and Variational MineGAN.}
    \label{tab:generation_times}
\end{table}

\begin{figure}[ht!]
\centering\includegraphics[width=\columnwidth]{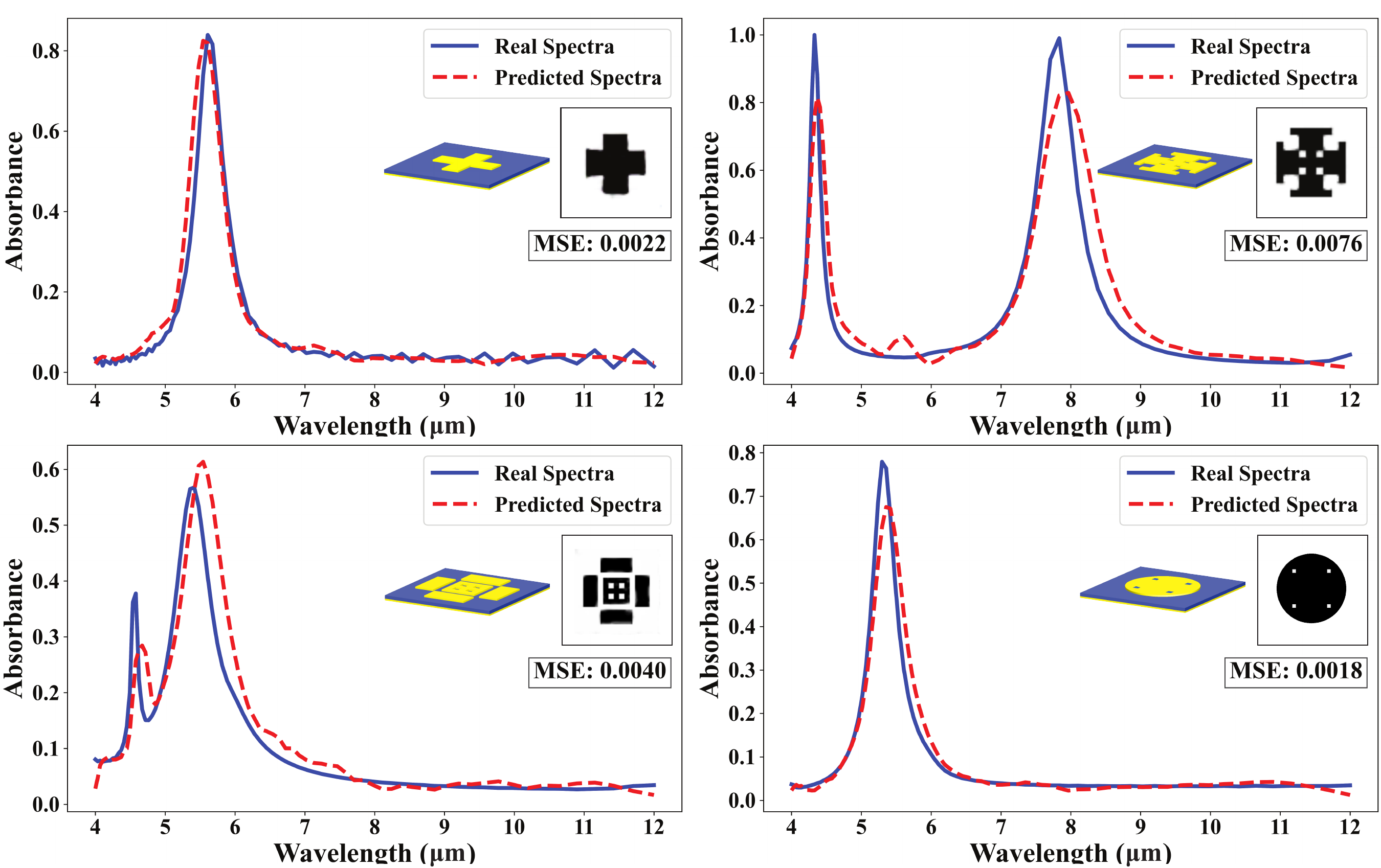}
\caption{Real and predicted spectra values for specific generated designs, with the corresponding design images, simulated versions and MSE values displayed as insets.}
\label{fig:Predicts}
\end{figure}

After generating a diverse set of images using Variational MineGAN, these images were passed through our trained predictive model to estimate their corresponding absorption spectra. This model, which was fine-tuned to minimize the prediction error between real spectral values and those predicted from design images, achieved a mean squared error (MSE) of \textbf{0.0046} on the validation dataset. As shown in Figure \ref{fig:Predicts}, the model approximates the spectral response for various designs pretty closely, with each subplot illustrating a comparison between the actual and predicted spectra for specific designs. The higher-quality design images from Variational MineGAN allow the predictor model to estimate the spectral response closely.

This step allows for an efficient and rapid exploration of the generated designs. By analyzing the predicted spectra, we can quickly identify the designs that best meet specific response criteria, such as a peak within a particular wavelength range or a peak higher than a specific value. After selecting the most promising design from the generated images, further validation can be performed through simulations, and other non-geometric parameters can be fine-tuned to optimize performance even further. This approach accelerates the iterative design process and enhances the practicality of using generative models for photonic design.

\section{Conclusion}
In this work, we introduced Variational MineGAN, a framework that enhances data-efficient transfer learning for generative AI-based design of nanophotonic structures. By leveraging a variational miner network to adapt the latent space, our approach alleviates the challenge of overfitting, commonly encountered in generative models, particularly in the case of small datasets. The architecture demonstrated significant improvements in performance over previous approaches, as measured by FID of 52.14 and IS of 3.59. This improvement in image quality not only facilitates the exploration of better designs but also allows for a more accurate estimation of spectral responses. Moreover, Variational MineGAN offers substantial time and resource efficiency during inference, enabling the rapid generation of high-quality designs compared to its predecessor, the MineGAN approach. The learning of target distribution space properties allows an efficient generation process, reducing time and computational cost, which makes it well-suited for large-scale applications. Although we employed the ProGAN architecture for our experiments, this framework can be extended to any generative network, making it a versatile tool for various design applications. Given that large datasets are often unavailable for many types of photonic design, Variational MineGAN provides an effective solution in such cases, empowering researchers to explore new possibilities in the design and optimization of intricate photonic structures.

\begin{backmatter}
\bmsection{Funding}
No external funding was available for this research.

\bmsection{Disclosures}
The authors declare no conflicts of interest.

\bmsection{Data Availability}
The code, along with the FDTD simulation data used for this work, are available at the following link: \url{https://github.com/Tarvirator/Variational-MineGAN}
\end{backmatter}

\bibliography{Variational_MineGan}

\end{document}